\documentstyle[12pt]{article}
\newcommand{\be}{\begin{equation}}
\newcommand{\ee}{\end{equation}}

\begin{document}
\begin{titlepage}
\sloppy
\thispagestyle{empty}

\mbox{}
\vspace*{\fill}
\begin{center}
{\LARGE\bf Sum Rules for Nucleon Spin Structure Functions } \\

\vspace{2mm}

\vspace{2mm}
\large
B.L.Ioffe 
\\
\vspace{2em}
{\it  Institute of Theoretical and\\
Experimental Physics
 \\
B.Cheremushkinskaya 25, 117259 Moscow, Russia}

e-mail: ioffe@vxdesy.desy.de, ioffe@vxitep.itep.ru
\end{center}
\vspace*{\fill}

\begin{abstract}
\noindent
The uncertainties in the perturbative and higher twist corrections to the
sum rules for $\Gamma_{p,n}$ are analyzed. The theoretical predictions for
$\Gamma_{p,n}$ are compared with the whole set of experimental data and the
restrictions on theoretical uncertainties was obtained. Suggestions for
future experiments are given.
\end{abstract}
\vspace*{\fill}

\end{titlepage}
\newpage
\section{The sum rules for $\Gamma_{p,n}$. Theoretical Status.}

I start with the consideration of the sum rules for the first
moments of the spin structure functions
\be
\Gamma_{p,n} (Q^2) = \int \limits_0 ^1 dx g_{1 p,n} (x, Q^2)
\ee
I will discuss the uncertainties in the theoretical predictions for
$\Gamma_{p,n}$ and compare the theoretical expectations with experimental
data. The aim of this consideration is to obtain from the experiment the
restrictions on the uncertainties in the theoretical description of the
problem.

Consider first the Bjorken sum rule [1]:
\be
\Gamma_p (Q^2) - \Gamma_n(Q^2) = \frac{1}{6} g_A [1 - \frac{\alpha_s(Q^2)}
{\pi} - 3.6 \left (\frac{\alpha_s(Q^2)}{\pi}\right )^2 - 20
\left (\frac{\alpha_s(Q^2)}{\pi}\right )^3] + \frac{b_{p-n}}{Q^2},
\ee
where $g_A$ is the axial $\beta$-decay coupling constant and the last term
in (2) represents the twist-4 correction. The perturbation QCD corrections
are known up to the third order $^{[2,3]}$ (there is also an estimate of the
fourth order term ${[^4]}$).
The
coefficients in (2) correspond to the number of flavours $N_f = 3$.

In what follows I will so often transfer the data to the standard reference
point $Q^2 = Q^2_0 = 10.5 GeV^2$ -- the mean value of $Q^2$, at which EMC and
SMC experiments were done.

Let us discuss the perturbative corrections in (2). Today there is a
serious discrepancy in the values of $\alpha_s$ found from different
experiments. The average value of $\alpha_s$, obtained at LEP is ${[^5]}$
\be
\alpha_s (m^2_Z) = 0.124 \pm 0.007
\ee
In two-loop approximation this value corresponds to the QCD parameter
$\Lambda_3$ for three flavours
\be
\Lambda_3 = 430 \pm 100 MeV
\ee
The data on the $\Upsilon \rightarrow$ hadrons decay give
$^{[5]}$ (the first $\alpha_s$ correction $^{[6]}$ is accounted)
\be
\alpha_s (m^2_b) = 0.178 \pm 0.010
\ee
from which it follows
\be
\Lambda_3 = 170 \pm 30 MeV
\ee
The small error in (6) is caused by the fact that the partial width
$\Gamma(\Upsilon \rightarrow 3g)$, from which (5) was determined, is
proportional to $\alpha^3_s$ and the $\alpha_3$ correction to it is small. A
strong contradiction of (4) and (6) is evident.

The overall fit $^{[7]}$ of the data of deep inelastic lepton-nucleon
scattering gives in the NLO approximation $\Lambda_3 = 250 MeV$ (the error
is not given). New data in the domain $m^2_Z$ indicate lower
$\alpha_s(m^2_Z)$  (SLD $^{[8]}: 0.118 \pm 0.013, 0.112 \pm 0.004$; OPAL
$^{[9]}$: $0.113 \pm 0.012$), but the data of AMY $^{[10]}$ on $e^+e^-$
annihilation at $\sqrt{s} = 57.3 GeV$ results in $\alpha_s(m^2_Z) = 0.120
\pm 0.005$. Finally, from $\tau$- decay it was obtained $^{[11]}$:
\be
\alpha_s(m^2_{\tau}) = 0.33 \pm 0.03
\ee
corresponding to
$$
\Lambda_3 = 380 \pm 60 MeV
$$
But the determination of $\alpha_s$ from $\tau$-decay can be criticized
$^{[12]}$ on the grounds that at such a low scale exponential terms in $q^2$
may persist in the domain of positive $q^2 = m^2_{\tau}$ besides the standard
power-like terms in $q^2$ accounted in the calculation.

In such a confusing situation I will consider two options -- of small and
large perturbative corrections. In the first case I will take $\Lambda_3 =
200 MeV$. Then
\be
\alpha_s(Q^2_0) = 0.180 \pm 0.010 ~~~~~~~~~~~~ \Gamma_p(Q^2_0) -
\Gamma_n(Q^2_0) = 0.194
\ee

In the second one $\Lambda_3 = 400 MeV$ and
\be
\alpha_s(Q^2_0) = 0.242 \pm 0.025 ~~~~~~~~~~~~~~~\Gamma_p(Q^2_0) -
- \Gamma_n(Q^2_0) = 0.187
\ee

The twist-4 contribution was disregarded in (8) and (9). There is also a
discrepancy in its value. The value of $b_{p-n}$ was determined in the QCD
sum rule approach by Balitsky, Braun and Koleshichenko (BBK) $^{[13]}$:
\be
b_{p-n} = -0.015 GeV^2
\ee
On the other hand, the model $^{[14]}$ based on connection $^{[15]}$ of the
Bjorken sum rule at large $Q^2$ with the Gerasimov, Drell, Hearn sum rule at
$Q^2 = 0$ $^{[16]}$ gives
\be
b_{p-n} = -0.15 GeV^2
\ee
In what follows I will consider (10) and (11) as two options which
correspond to small (S) and large (L) twist-4 corrections. I will
discuss both approaches of determination of higher twist corrections in more
details below. From my point of view, no one of these approaches is
completely reliable and I will use them in comparison with experiments only
as reference points.

I turn now to the sum rules for $\Gamma_p$ and $\Gamma_n$:
\be
\Gamma_{p,n}(Q^2) = \frac{1}{12} \left \{\left [1 - \frac{\alpha_s}{\pi} -
3.6 (\frac{\alpha_s}{\pi})^2 - 20 (\frac{\alpha_s}{\pi})^3\right ] (\pm g_a
+ \frac{1}{3} a_8) + \right.
\ee
$$\left.\frac{4}{3} \left [1 -
\frac{\alpha_s}{3\pi} - 0.55 (\frac{\alpha_s}{\pi})^2\right ] \Sigma \right
\} - \frac{N_f}{18 \pi} \alpha_s (Q^2) \Delta g (Q^2) + \frac {b_{p,n}}{Q^2}
$$

According to the current algebra $g_A, a_8$ and $\Sigma$ are determined by
the proton (or neutron) matrix elements of flavour octet and singlet axial
currents
\be
-2m s_{\mu} a_8 = <p, s \mid j^{(8)}_{5 \mu} \mid p,s > ~~~~~ -2ms_{\mu}
\Sigma = <p, s \mid j^{(0)}_{5 \mu} \mid p,s>,
\ee
$$ -2ms_{\mu} g_A = <p, s \mid j^{(3)}_{5 \mu} \mid p,s >
{}~~~~~~~~~~~(13')$$
where
$$j^{(8)}_{5 \mu} = \bar{u} \gamma_{\mu} \gamma_5 u + \bar{d}
\gamma_{\mu} \gamma_5 d - 2 \bar{s} \gamma_{\mu} \gamma_5 s,~~~~~ j^{(3)}_{5
\mu} = \bar{u} \gamma_{\mu} \gamma_5 u - \bar{d} \gamma_{\mu} \gamma_5 d$$

$$
j^{(0)}_{\mu} = \bar{u} \gamma_{\mu} \gamma_5 u + \bar{d}
\gamma_{\mu} \gamma_5 d + \bar{s} \gamma_{\mu} \gamma_5 s
$$
In the parton model $g_A,a_8$ and $\Sigma$ are equal to
\be
g_A = \Delta u - \Delta d ~~~~~~~~ a_8 = \Delta u + \Delta d - 2\Delta S
{}~~~~~~~\Sigma = \Delta u + \Delta d + \Delta s,
\ee
where
\be
\Delta q = \int \limits _{0} ^{1} \left [q _+(x) - q_-(x)\right ]~~~~~~~~~~
q = u,d,s
\ee
and $q_{\pm}$ are the quark distributions with the spin parallel
(antiparallel) to the proton spin, which is supposed to be longitudinal
(along the beam). $\Delta g$ in (12) has the similar meaning, but for
gluons, as  $\Delta q$ for quarks
\be
\Delta g(Q^2) = \int \limits _{0} ^{1} dx \left [ g_+ (x,Q^2) - g_- (x,
Q^2)\right ]
\ee
Unlike $g_A, a_8$ and $\Sigma$ which in the approximation used above are $Q^2$
independent and have zero anomalous dimensions, $\Delta g$ anomalous
dimension is equal to --1. This means that
\be
\Delta
g(Q^2)_{Q^2 \rightarrow \infty} \simeq cln Q^2
\ee
The conservation of the projection of the angular
momentum can be written as
\be
\frac{1}{2} \Sigma + \Delta g (Q^2) + L_z
(Q^2) = \frac{1}{2}
\ee
where $L_z$ has the meaning of the orbital momentum
of quarks and gluons.  As follows from (17),(18) at high $Q^2 ~~ L_z(Q^2)$
must compensate $\Delta g(Q^2)$, $L_z(Q^2) \approx -c ln Q^2$. This means
that the quark model, where all quarks are in $S$-state, failed with $Q^2$
increasing.

Now about the numerical values of the constants $g_A$, $a_8$ and $\Sigma$
entering eq.(12). $g_A$ is known with a very good accuracy, $g_A = 1.257 \pm
0.003$  $^{[5]}$.

Under assumption of SU(3) flavour symmetry in baryon decays $a_8$ is equal
to
\be
a_8 = 3F - D = 0.59 \pm 0.02
\ee
where $F$ and $D$ are axial coupling constants of baryon $\beta$-decays in
SU(3) symmetry and the numerical value in the r.h.s. of (19) follows
from the best fit to the data $^{[17]}$. The combination $3F-D$ can be found
from any pair of baryon $\beta$-decays. The comparison of the values of
$3F-D$, obtained in this way shows, that the spread is rather narrow
$^{[18]}$, $\mid \delta a_8 \mid \leq 0.05$ and at least $a_8 > 0.50$. This
may be an argument in the favour that SU(3) violation is not large
here.(See,however, $^{[19]}$). It should be mentioned that at fixed
$\Gamma_{p,n}$ the uncertainty in $a_8$ only slightly influences the most
interesting quantities $\Sigma$ and $\Delta s$.  As follows from (12) and
(14)
\be
\delta \Sigma = - \delta (\Delta s) = \frac{1}{4} \delta a_8
\ee
$a_8$ was also determined by the QCD sum rule method $^{[20]}$. In this
approach no SU(3) flavour symmetry was assumed and the result
\be
a_8 = 0.5 \pm 0.2
\ee
is in agreement with (19), although the error is large.

What can be said theoretically about $\Sigma$? In their famous paper
$^{[21]}$ Ellis and Jaffe assumed that the strange sea in the nucleon is
nonpolarized, $\Delta s = 0$. Then
\be
\Sigma \approx a_8 \approx 0.60
\ee
This number is in a contradiction with the experimental data
pointing to smaller values of $\Sigma$. On the other hand, Brodsky, Ellis
and Karliner $^{[22]}$ had demonstrated that in the Skyrme model at large
number of colours $N_c$, $\Sigma \sim 1/N_c$ and is small. From my point of
view this argument is not very convincing: the Skyrme model may be a good
model for description of nucleon periphery, but not for the internal part of
the nucleon determining the value of $\Sigma$ (see also $^{[23]}$).

An attempt to calculate $\Sigma$ in the QCD sum rule approach in the way
similar to the $a_8$ determination $^{[20]}$, failed. It was found
$^{[24]}$, that the operator product expansion (OPE) breaks down in case of
the polarization operator (and/or NNA vertex) for singlet axial current at
the scale $\sim 1 GeV$. (The anomaly was properly accounted in the
calculation). This is indicative of that the higher dimension operators
(instantons?) are of importance in this problem. The physical consequence is
that one may expect a violation of the Okubo-Zweig-Iizuka rule in the nonet
of axial mesons. The similar trouble, perhaps, faces attempts to determine
$\Sigma$ using the so called "$U(1)$ Goldberger-Treiman relation" (for a
review see $^{[25]}$). So, at this stage, the only way to find $\Sigma$ is
from an experiment, exploiting eq.(12).

I dwell now on the determination of higher twist corrections. The twist-4
correction was calculated by BBK $^{[13]}$, using the QCD sum rule method
for the vertex function in the external field $^{[26,27]}$. The result for
$b_{p-n}$ was given in (10). For $b_{p+n}$ it was obtained
\be
b_{p+n} = -0.022 GeV^2
\ee
This result, however, cannot be considered as reliable for the following
reasons:

1. BBK use the same hypothesis as Ellis and Jaffe did, i.e. assume that
$s$-quarks do not contribute to the spin structure functions and instead of
singlet (in flavour) operator consider the octet one.

2. When determining the induced by external field vacuum condensates, which
are very important in such calculations (see $^{[26]}$), they saturate the
corresponding sum rule by $\eta$-meson contribution, what is wrong. (Even
the saturation by $\eta'$-meson would not be correct, since $\eta'$ is
not a Goldstone).

3. One may expect that in the same way as in the calculation of $\Sigma$ by
the QCD sum rule, in this problem the OPE series diverge at the scale $\sim 1
GeV$, where the BBK calculation proceeds.

Even in the case of the Bjorken sum rule, where the mentioned above problems
are absent, the value $b_{p-n}$ (10) is questionable. $b_{p-n}$ was
calculated by BBK using the OPE for the vertex function and accounting few
terms in OPE with operator dimensions from 0 to 8. But the final result
comes almost entirely from the last term of OPE -- the operator of dimension
8. (There are other drawbacks in these calculations -- see $^{[28]}$). It is
clear that such situation is unsatisfactory. Recently, Oganesian $^{[29]}$
had calculated the next term in OPE for $b_{p-n}$ determination -- the
dimension 10 term and found that by absolute value it is equal to (10), but
has opposite sign, so the total result is zero. Of course, we cannot believe
in this statement either: it means only that the results of the calculations
are unstable and the value (10)  characterizes the answer by the order of
magnitude only.

The other way to determine higher twist corrections suggested in $^{[14]}$
is based on the connection of $\Gamma_{p,n}(Q^2)$ at high $Q^2$ with the
Gerasimov-Drell-Hearn $^{[16]}$ (GDH) sum rule at $Q^2 = 0$. The idea is the
following. For the real photon there is only the spin-dependent
photon-nucleon scattering amplitude $S_1(\nu)$, for which we can write an
unsubtracted dispersion relation
\be
S_1(\nu) = 4 \int \limits _{0} ^{\infty}~\nu' d\nu'~ \frac{G_1(\nu',
0)}{\nu^{'~2} - \nu^2}
\ee
Consider
the limit $\nu \rightarrow 0$ in (24). According to the F.Low theorem the
terms proportional to $\nu^0$ and $\nu^1$ in the expansion in powers of
$\nu$ of the photon-nucleon scattering amplitude are
expressed via static characteristics of nucleon.

The calculation gives
\be
S_1(\nu)_{\nu \rightarrow 0} = - \kappa^2,
\ee
where $\kappa$ is the nucleon anomalous magnetic moment: $\kappa_p = 1.79,
\kappa_n = -1.91$.

{}From (24), (25)  the GDH sum rule follows:
\be
\int \limits _{0} ^{\infty} \frac {d \nu}{\nu}~ G_1(\nu, 0) = - \frac{1}{4}
\kappa^2
\ee
Till now only indirect check of (26)was performed,
where in the l.h.s the parameters of resonances, obtained from the
$\pi N$  scattering phase analysis, where substituted. In this way with
resonances up to 1.8 GeV it was obtained $^{[30]}$  (see also the second
reference in $^{[14]}$ and $^{[31]}$)
\be
\begin{array}{lll}
& ~~~~\mbox{l.h.s of (26)} & ~~~~~~~\mbox{r.h.s. of (26)}\\
\mbox{proton}& ~~~~-1.03 & ~~~~~~~ -0.803 \\
\mbox{neutron} & ~~~~-0.83 & ~~~~~~~-0.913\\
\end{array}
\ee
The l.h.s.  and the r.h.s. of (26) are not in a good agreement -- a
nonresonant contribution is needed. The direct check of the GDH sum rule
would be very desirable!

An important remark: the forward spin dependent photon--nucleon scattering
amplitude has no nucleon pole. This means that there is no nucleon
contribution in the l.h.s. of GDH sum rule -- all contributions come from
excited states: the GDH sum rule is very nontrivial.

In order to connect the GDH sum rule with $\Gamma_{p,n}(Q^2)$  consider the
integrals $^{[15]}$
\be
I_{p,n}(Q^2) = \int^{\infty}_{Q^2/2}~\frac{d\nu}{\nu}G_{1,p,n}(\nu,Q^2)
\ee
It is easy to see that at large $Q^2$
\be
I_{p,n}(Q^2) = \frac{2m^2}{Q^2}~\Gamma_{p,n}(Q^2)
\ee
and at $Q^2=0$ (28) reduces to the GDH sum rule. For proton $I_p$ is
negative at $Q^2=0$  and positive at large $Q^2$, what indicates
to large nonperturbative corrections. In $^{[14]}$ the VDM based
interpolation model was suggested, describing $I_{p,n}(Q^2)$ in the whole
domain of $Q^2$.  The model was improved in the second Refs. 14, where the
contributions of baryonic resonances up to $W=1.8$GeV, taken from
experiment, where accounted. The model has no free parameters, besides the
vector meson mass, for which the value $\mu^2_V=0.6 GeV^2$ was chosen. Using
this model it is possible to calculate the higher twist
contributions in (2),(12). The results are presented in Table 1, as the
ratio of asymptotic $\Gamma^{as}$ with power corrections excluded to the
experimentally measurable $\Gamma$  at given $Q^2$ $^{[18]}$.

\vspace{5mm}
\centerline{\bf Table 1.}

\vspace{3mm}
\centerline{\underline{Higher twist corrections in GDH sum rule + VDM
inspired model.}}

\vspace{3mm}
\begin{center}
\begin{tabular}{|l|c|c|c|c|}\hline
$Q^2(GeV^2)$  & ~~~~2 & ~~3 & 5 ~~& 10 \\ \hline
$\Gamma^{as}_p/\Gamma_p$ & ~~~~1.44 & ~~1.29 & ~~1.18 & ~~1.08 \\
$\Gamma^{as}_n/\Gamma_n$ & ~~~~1.30 & ~~1.20 & ~~1.13 & ~~1.06 \\
$\Gamma^{as}_{p-n}/\Gamma_{p-n}$ &  ~~~~1.45 & ~~1.29 & ~~1.18 &~ 1.08 \\
$\Gamma^{as}_{p+n}/\Gamma_{p+n}$ &  ~~~~1.47 & ~~1.31 & ~~1.19 & ~~1.08 \\
\hline
\end{tabular}
\end{center}

\vspace{5mm}
The power corrections, given in Table 1 are essentially larger (except for
the case of neutron), than the values (10),(23) found in $^{[13]}$.
It must be mentioned that the accuracy of the model in the domain
of intermediate $Q^2$, where it is exploited, is not completely certain.

\vspace{5mm}
\section{Comparison with experiment.}

When comparing the sum rules (2),(12) with experiment I consider two
limiting variants of perturbative corrections: small with $\Lambda_3=200$MeV
and large with $\Lambda_3=400$MeV. ($\alpha_s$ is computed in 2- loop
approximation, it is assumed that the number of flavours $N_f=3$). For
higher twist correction I also consider two limiting options: small (S),
given by (10),(23) and large (L), determined by the data of Table 1. The
contribution of gluons $\Delta g(Q^2)$ in (12) will be found in the
following way. Let us assume, that at 1 GeV the quark model is valid and
$L_z(1$GeV$^2)=0$  in eq.(18). Taking $\Sigma$=0.3, what is a reasonable
average of the data, we have from (18)
\be
\Delta g (1GeV^2) = 0.35
\ee
The $Q^2$  dependence of $\Delta g$ can be found from the evolution equation
$^{[32]}$
$$ \Delta g(Q^2) = \frac{\alpha_s(\mu^2)}{\alpha_s(Q^2)}\left \{ 1 +
\frac{2N_f}{b\pi}\left [\alpha_s(Q^2) - \alpha_s(\mu^2)\right ] \right \}
\Delta g(\mu^2) + $$
\be
+ \frac{4}{b}\left [ \frac{\alpha_s(\mu^2)}{\alpha_s(Q^2)} - 1 \right ]
\Sigma(\mu^2),
\ee
where $b=11-(2/3)N_f = 9,~ \mu^2 = 1 GeV^2$  and $\Sigma(1 GeV^2) \approx
0.3$. As the calculation shows, the change of scale at which the quark model
is assumed to work (say $0.5 GeV^2$ instead of $1 GeV^2$) or the of use
slightly different $\Sigma(\mu^2)$ in (31)  only weakly influence the
results for $\Sigma$ and $\Delta s$, obtained from experimental data.

I consider the following experimental data (Table 2).

\vspace{5mm}
\centerline{\bf Table 2.}

\vspace{2mm}
\centerline{\underline{The experimental data on $\Gamma_{p,n}$.}}

\vspace{2mm}
\begin{center}
\begin{tabular}{|l|c|l|c|}\hline
$\mbox{Experimental}$  & The target & ~~~~~~~~~~$\Gamma_(p,n)$ &
 \mbox{Mean} $\overline{Q^2} (GeV^2)$ \\
\mbox{group} & & & \\ \hline
EMC [33] & $p$ & $\Gamma_p = 0.126 \pm 0.010 \pm 0.015$ & 10.7 \\
SMC [34] & $p$ & $\Gamma_p = 0.136 \pm 0.011 \pm 0.011$ & 10.5 \\
E143 [35] & $p$ & $\Gamma_p = 0.127 \pm 0.004 \pm 0.010$ & 3 \\
E142 [36] & $He^3$ & $\Gamma_n = -0.022 \pm 0.011$ & 2 \\
SMC [37] & $d$ & $\Gamma_d = 0.034 \pm 0.009 \pm 0.06$ & 10 \\
 & & $\Gamma_p + \Gamma_n = 0.073 \pm 0.022$ & 10 \\
 E143 [38] & $d$ & $\Gamma_d = 0.042 \pm 0.003 \pm 0.004$ & 3\\
 & & $\Gamma_p + \Gamma_n = 0.0908 \pm 0.006 \pm 0.008$& 3 \\
\hline
\end{tabular}
\end{center}

\newpage
\vspace{5mm}
In
comparison with experiment the perturbative and higher twist corrections, as
well as $\Delta g (Q^2)$ contributions are calculated for $\overline
{Q^2}$. Experimentally, $Q^2$ are different in different $x$-bins (higher
$Q^2$ at larger $x$). This effect is not accounted in the calculation. The
ratio of $\alpha^3_s$ term to $\alpha^2_s$ term in perturbative corrections
is of order of 1 at $\Lambda_3 = 400 MeV$ and $Q^2 \approx 2-3 GeV^2$ (as
well as $\alpha^4_s/\alpha^3_s$ estimate). For this reason we introduce in
these cases an additional error equal to the $\alpha^3_s$. The values of
$\Sigma$ and $\Delta s$ calculated from comparison of experimental data with
eq.12 are shown in Table 3.(The errors are summed in quadrature).

\vspace{5mm}
\centerline{\bf Table 3.}
\vspace{2mm}
\centerline{\underline{Determination of $\Sigma$ and $\Delta s$ from
experimental data}}

\vspace{2mm}
\begin{center}
\begin{tabular}{|c|c|c|c|c|}\hline
Experiment & $\Lambda_3$     & High & $\Sigma$ & $\Delta s$\\
           &~~MeV ~~         & twist &       &            \\ \hline
           & 200            & S    & $0.21 \pm 0.17$ & $-0.13 \pm 0.06$ \\
           \cline{2-4}
 EMC       & 400            & S    & $0.29 \pm 0.17$ & $-0.10 \pm 0.06$ \\
 \cline{2-5}
    p      & 200            & L    &$0.285 \pm 0.17$ & $-0.10 \pm 0.06$\\
 \cline{2-5}
           & 400            & L    & $0.37 \pm 0.17$ & $-0.07 \pm 0.06$ \\
           \hline
           & 200            & S    & $0.30 \pm 0.14$ & $-0.10 \pm 0.05$ \\
           \cline{2-5}
 SMC       & 400            & S    & $0.39 \pm 0.14$ & $-0.07 \pm 0.05$ \\
 \cline{2-5}
  p       & 200            & L    & $0.39 \pm 0.14$ & $-0.07 \pm 0.05$\\
 \cline{2-5}
          & 400            & L    & $0.47 \pm 0.14$ & $-0.04 \pm 0.05$ \\
 \hline
           & 200            & S    & $0.28 \pm 0.10$ & $-0.10 \pm 0.03$  \\
           \cline{2-5}
 E143      & 400            & S    & $0.42 \pm 0.10$ & $-0.06 \pm 0.03$  \\
 \cline{2-5}
    p      & 200            & L    & $0.57 \pm 0.10$ & $-0.006 \pm 0.03$ \\
 \cline{2-5}
           & 400            & L    & $0.71 \pm 0.10$ & $ 0.04 \pm 0.03$ \\
           \hline
 E142      & 200            & S    & $0.60 \pm 0.12$ & $0.003 \pm 0.04$ \\
           \cline{2-5}
  n        & 400            & S    & $0.57 \pm 0.12$ & $-0.005 \pm 0.04$ \\
  \cline{2-5}
 ($^3He$)    & 200            & L    & $0.64 \pm 0.12$ & $0.016 \pm 0.04$ \\
 \cline{2-5}
           & 400            & L    & $0.61 \pm 0.12$ & $0.008 \pm 0.04$ \\
           \hline
           & 200            & S    & $0.27 \pm 0.10$ & $-0.11 \pm 0.03$ \\
           \cline{2-5}
 SMC       & 400            & S    & $0.33 \pm 0.10$ & $-0.09 \pm 0.03$ \\
 \cline{2-5}
 d         & 200            & L    & $0.29 \pm 0.10$ & $-0.10 \pm 0.03$\\
 \cline{2-5}
           & 400            & L    & $0.34 \pm 0.10$ & $-0.08 \pm 0.03$ \\
           \hline
           & 200            & S    & $0.37 \pm 0.06$ & $-0.07 \pm 0.02$ \\
           \cline{2-5}
 E143      & 400            & S    & $0.44 \pm 0.06$ & $-0.05 \pm 0.02$ \\
 \cline{2-5}
     d     & 200            & L    & $0.48 \pm 0.06$ & $-0.04 \pm 0.02$\\
 \cline{2-5}
           & 400            & L    & $0.54 \pm 0.06$ & $-0.015 \pm 0.02$ \\
           \hline
\end{tabular}
\end{center}
\vspace{3mm}
Remark: the contribution to $\Sigma$ of the term proportional to $\Delta g$
is approximately equal to 0.06 in the case of $\Lambda_3 = 200 MeV$ and 0.11
in the case of $\Lambda_3 = 400 MeV$.

If we assume that all the analysed above experiments are correct in the
limits of their quoted errors (or, may be, 1.5 st.deviations), then
requiring  for the results for $\Sigma$ and $\Delta s$ from various
experiments to be consistent, we may reject some theoretical possibilities.
A look at the Table 3 shows that the variant $\Lambda_3 = 400 MeV, L$ (a
contradiction of E143, p  and SMC, d results for $\Sigma$) and, less
certain, the variant $\Lambda_3 = 200 MeV, S$ (a contradiction of E142, n and
SMC, d) may be rejected.

\newpage

Consider now the Bjorken sum rule. For comparison with theory I choose
combinations of the SMC data - on proton and deuteron, the E143 data - on
proton and deuteron and the E143 data on proton and the E142 on neutron
($^3He$). The results of the comparison of the experimental data with the
theory are given in Table 4.

\vspace{5mm}
\centerline{\bf Table 4.}
\vspace{2mm}
\centerline{\underline{Comparison of the experimental data with the Bjorken
sum rule}}

\vspace{2mm}
\begin{center}
\begin{tabular}{|c|c|c|c|c|} \hline
Combination & $(\Gamma_p - \Gamma_n)_{exper.}$ & $\Lambda_3(MeV)$ & High &
$(\Gamma_p - \Gamma_n)_{th}$ \\
of experiments &                             &                   & twist &
\\ \hline
            &                                &    200            & S & 0.193
            \\ \cline{3-5}
  SMC, p    & $0.199 \pm 0.038$              &    400            & S & 0.186
  \\ \cline{3-5}
  SMC, d    &                                &    200            & L & 0.180
  \\ \cline{3-5}
            &                                &    400            & L & 0.173
            \\ \hline
  E143,p          &                          &    200            & S & 0.182
            \\ \cline{3-5}
            & $0.163 \pm 0.010 \pm 0.016$     &    400          & S & 0.168
            \\ \cline{3-5}
  E143,d    &                                &    200            & L & 0.145
            \\ \cline{3-5}

            &                                &    400            & L & 0.134
  \\ \hline
  E143,p &                                   & 200               & S & 0.182
  \\ \cline{3-5}
  E142 n($^3He$) & $0.147 \pm 0.015$          & 400               & S & 0.168
  \\ \cline{3-5}
  recalculated to &                            & 200               & L &
  0.145 \\ \cline{3-5} $\bar{Q}^2 = 3 GeV^2$ &                    & 400
  & L & 0.134 \\ \hline
  \end{tabular}
  \end{center}
  \vspace{3mm}
  From Table 4
we see again some indications for rejection of variants $\Lambda_3 = 400
MeV$, L and, less certain, $\Lambda_3 = 200 MeV$, S.(In the first case there
is a contradiction of the theory with the E143,p and d data, in the second -
with the E143,p , E142, n data).

At existing
experimental accuracy it is impossible to choose from the data the true
values of $\Lambda_3$ and twist-4 correction .
My personal preference is to the variant $\Lambda_3 = 200 MeV$ and to the
value of twist-4 corrections 3 times smaller than given by the GDH sum rule
+ VDM inspired model and, correspondingly, $b_p = 0.04$ in (12), i.e., 2.2
times larger than the BBK result. The argument in the favour of such choice
is that at larger $\Lambda_3$ there will arise many contradictions with the
description of hadronic properties in the framework of the QCD sum rules.
The recent preliminary SLAC data $^{[39]}$ on $g_1(x, Q^2)~ Q^2$-dependence
indicate that $\Gamma^{as}_p / \Gamma_p = 1 + c_p/Q^2, c_p = 0.25 \pm 0.15$
what is compatible with the estimate above. In this case all experimental
data except for E142,n, are in a good agreement with one another and the
values of $\Sigma$ and $\Delta s$ averaged over all experiments, except for
E142,n are
\be
\overline{\Sigma} = 0.35 \pm 0.05 ~~~~~~~~~~~ \Delta s =
-0.08 \pm 0.02
\ee
(see also $^{[42]}$ where the values close to (32) were
obtained).

The values of $\Sigma$ and $\Delta s$ obtained from the E142,n experiment at
such a choice of $\Lambda_3$ and twist-4 corrections, are different:
\be
\Sigma = 0.61 \pm 0.12 ~~~~~~~~~~ \Delta s = 0.01 \pm 0.04
\ee
It is impossible to compete (32),(33) by any choice of $\Lambda_3$ and
higher twist correction. Perhaps, this difference is caused by inaccounted
systematic errors in the E142 experiment.

\newpage

A remarkable feature of the result (32) (as well as of the data in Table 3)
is the large value of $\mid \Delta s \mid$ - the part of the proton spin
projection carried by strange quarks. This value may be compared with the
part of the proton momentum carried by strange quarks
\be
V^s_2 = \int~ dx ~ x \left [ s_+(x) + s_-(x)\right ] = 0.026 \pm
0.006~^{[40]}, ~~~~ 0.040 \pm 0.005~^{[41]}
\ee
The much larger value of
$\mid \Delta s \mid$ in comparison with $V^s_2$ contradicts the standard
parametrization
$$s_+(x) + s_-(x) = A~x^{-\alpha}~ (1 -
x)^{\beta}$$
\be
s_+(x) - s_-(x) = B~x^{-\gamma}~(1-x)^{\beta}
\ee
and requirement of
positiveness of $s_+$ and $s_-$, if $\alpha \approx 1$(pomeron intercept) and
$\gamma \leq 0$ ($a_1$ intercept). Large $\mid \Delta s \mid$ and small
$V^s_2$ means that the transitions $\bar{s} s \rightarrow \bar{u} u +
\bar{d} d$ are allowed in the case of the operator $j_{\mu 5}$ and are
suppressed in the case of the quark energy-momentum tensor operator
$\Theta_{\mu \nu}$, corresponding to the matrix element $V_2$. Such
situation can be due to nonperturbative effects and the instanton
mechanism for its explanation was suggested $^{[43]}$. Improvement of
experimental accuracy is necessary in order to be sure that the inequality
$\mid \Delta s \mid \gg V^s_2$ indeed takes place.

\vspace{5mm}
\section{The calculations of the polarized structure functions by the QCD
sum rule method}

\vspace{2mm}

I mention here only some basing points of the calculations referring
for details to $^{[44]}$. The calculation was performed on the basis of OPE
with the account of the unit operator and of the square of quark condensate
$\sim \alpha_s <0 \mid \bar{\psi} \psi \mid 0 >^2$. It was shown that for
the bare quark loop the Bjorken and Burkhardt-Cottingham sum rules are
fulfilled. It was also found that for the function $g_1(x)$ the results are
reliable in a rather narrow domain of intermediate $x$ : $0.5 \leq x \leq
0.7$. In this domain the contribution of $u$-quarks $g^u_1$ is much larger
than d-quarks, $g^u_1 \gg g^d_1$. Therefore, $g_1 \approx (4/9)~g^u_1$.
$g^u_1$ was calculated and for the mean value of $g_1$ in this interval it
was obtained (at $Q^2 \sim 5-10 GeV^2$):
\be
\bar{g}_1 (0.5 < x < 0.7)
= 0.05 \pm 50\%
\ee
The large uncertainty in (36) results from the large
contribution of nonleading term in OPE and from large background at the
phenomenological side of the QCD sum rule. The E143 proton $^{[35]}$ and
deuteron $^{[38]}$ data (the latter under assumption that $g^n_1$ is small
in this interval of $x$) give roughly the same values:
\be
\bar{g}_1~
(0.5 \leq x \leq 0.7) = 0.08 \pm 0.02
\ee
This value is in a good agreement
with the SMC result $^{[34]}$
\be
\bar{g}_1~ (0.4 \leq x \leq 0.7) =
0.08 \pm 0.02 \pm 0.01
\ee
and with our theoretical expectation (36).

For  the case of the structure function $g_2$ only $g^u_2$ can be calculated
at $0.5 < x < 0.8$, the calculation of $g^d_2$ fails because of large
contribution of nonleading terms in OPE. If we assume that like in the case
of $g_1,~\mid g^d_2 \mid \ll \mid g^u_2 \mid$,
\be
g_2 (0.5 < x < 0.8) = -0.05 \pm 50\%
\ee
The E143 data $^{[45]}$ in this interval of $x$ are:
\be
g_2(0.5 < x < 0.8) = -0.037 \pm 0.020 \pm 0.003
\ee
in a good agreement with (39).

\vspace{5mm}
\section{What would be desirable to do experimentally in the near future?}

\vspace{2mm}
1. To increase the accuracy by 2-3 times.\\
2. To study the $Q^2$-dependence (in separate bins in $x$).\\
3. To go to higher $Q^2$ (at HERA). The experiment at higher $Q^2$ will be
informative if only its accuracy will not be worse than the existing ones.\\
4. To have better data in the domain of small $x$. (Probably, this also can
be done only at HERA).\\
5. Measurements of two-jets events in polarized deep-inelastic scattering,
which will give information about $\Delta g$.\\
6. To have
more data on $s$-quark distribution in nonpolarized nucleon, particularly at
$x < 0.1$.\\ 7. To have better data for $g_2$.\\ 8. To perform a direct
check of the GDH sum rule.\\

\vspace{3mm}
\centerline{\bf Acknowledgement}

\vspace{3mm}
I am very indebted to B.Wiik, A.Levy and W.Buchm\"uller for their hospitality
at DESY, where this work was finished. This investigation was supported in
part by the International Science Foundation Grant M9H300 and by the
International Association for the Promotion of Cooperation with Scientists
from Independent States of the Former Union Grant INTAS-93-0283.

\end{document}